\documentclass[manuscript,screen]{acmart}

\usepackage{svg}
\usepackage{multirow}
\usepackage{booktabs}
\usepackage{colortbl}
\newtheorem{definition}{Definition}
\usepackage[super]{nth}

\AtBeginDocument{%
  \providecommand\BibTeX{{%
    \normalfont B\kern-0.5em{\scshape i\kern-0.25em b}\kern-0.8em\TeX}}}

\setcopyright{acmcopyright}
\copyrightyear{2018}
\acmYear{2018}
\acmDOI{10.1145/1122445.1122456}

\acmConference[Woodstock '18]{Woodstock '18: ACM Symposium on Neural
  Gaze Detection}{June 03--05, 2018}{Woodstock, NY}
\acmBooktitle{Woodstock '18: ACM Symposium on Neural Gaze Detection,
  June 03--05, 2018, Woodstock, NY}
\acmPrice{15.00}
\acmISBN{978-1-4503-XXXX-X/18/06}

\begin{document}

\title{Group Validation in Recommender Systems: Framework for Multi-layer Performance Evaluation}





%

%
%


\author{Wissam Al Jurdi}
\email{wissam.aljurdi@st.ul.edu.lb}
\email{wissam.al_jurdi@univ-fcomte.fr}
\orcid{0000-0001-9497-0515}
\affiliation{%
  \institution{University Bourgogne Franche-Comté, FEMTO-ST Institute, CNRS, Montbéliard, France | LaRRIS, Faculty of Sciences, Lebanese University}
  \city{Beirut}
  \country{Lebanon}}

\author{Jacques Bou Abdo}
\orcid{0000-0002-3482-9154}
\affiliation{%
  \institution{University of Nebraska at Kearney}
  \city{Kearney}
  \state{Nebraska}
  \country{USA}}
\email{bouabdoj@unk.edu}

\author{Jacques Demerjian}
\orcid{0000-0001-9798-8390}
\affiliation{%
  \institution{LaRRIS, Faculty of Sciences, Lebanese University}
  \city{Fanar}
  \country{Lebanon}}
\email{jaques.demerjian@ul.edu.lb}

\author{Abdallah Makhoul}
\orcid{0000-0003-0485-097X}
\affiliation{%
 \institution{University Bourgogne Franche-Comté, FEMTO-ST Institute, CNRS, Montbéliard}
 \country{France}}
\email{abdallah.makhoul@univ-fcomte.fr}

\renewcommand{\shortauthors}{Al Jurdi, et al.}

\begin{abstract}
Interpreting the performance results of models that attempt to realize user behavior in platforms that employ recommenders is a big challenge that researchers and practitioners continue to face. Although current evaluation tools possess the capacity to provide solid general overview of a system's performance, they still lack consistency and effectiveness in their use as evident in most recent studies on the topic. Current traditional assessment techniques tend to fail to detect variations that could occur on smaller subsets of the data and lack the ability to explain how such variations affect the overall performance. In this article, we focus on the concept of data clustering for evaluation in recommenders and apply a neighborhood assessment method for the datasets of recommender system applications. This new method, named neighborhood-based evaluation, aids in better understanding critical performance variations in more compact subsets of the system to help spot weaknesses where such variations generally go unnoticed with conventional metrics and are typically averaged out. This new modular evaluation layer complements the existing assessment mechanisms and provides the possibility of several applications to the recommender ecosystem such as model evolution tests, fraud/attack detection and a possibility for hosting a hybrid model setup.
\end{abstract}

\ccsdesc[500]{Information systems~Recommender systems}

\keywords{recommender systems, offline evaluation, model validation, data clustering}

\maketitle

\section{Introduction}

Recommendation systems are designed to effectively analyze information in datasets to generate predictions of what can potentially be of interest to users, which is done through utilizing available features such as their search and interaction data. While it is a relatively easy task to accomplish with the current abundance of models and research on the topic, evaluating the effectiveness and testing performance as models evolve with their underlying data remains, to date, a very tough challenge \cite{sun2020we,ricci2021challenges,ovaisi2022rgrecsys,macdonald2021simpson,bellogin2021improving,tamm2021quality}. There have been several approaches to tackle the evaluation challenge in recommender systems. These approaches include the Simpson's paradox \cite{macdonald2021simpson}, benchmarking toolkits \cite{sun2020we,anelli2021elliot,zhao2021recbole}, dataset-oriented design \cite{chin2022datasets} and metric selection criteria \cite{tamm2021quality,han2021and}. Although these topics tend to be diverse in nature, almost all of them revolve around one common idea: the tools for assessing models still require major improvements. In short, there hasn't been enough attempts to combine what is known about the evaluation of recommender systems, nor to systematically define the implications of evaluating recommenders for different tasks and under different contexts.

The topic of evaluation challenges in recommender systems is not recent and in fact dates back to 2004. At the time, Herlocker et al. \cite{macdonald2021simpson} abstracted all the factors that were considered in evaluating the effectiveness of a recommender model and proved the presence of potential biases in most reported evaluation results. Back then, the study highlighted the pitfalls that researchers ought to avoid when attempting to implement recommendation solutions, especially during the phase of trained models' effectiveness evaluation. Some of the issues are: undefined user goals for a particular recommender system, evaluation of models on inappropriate datasets, incorrect use of evaluation metrics that's mostly isn't based on the user goals for the system, etc. Surprisingly, such factors are still causing major issues to date as evident by one of the very recent comprehensive studies on noise and evaluation, Al Jurdi et al. \cite{jurdi2021critique}. In this research of \cite{jurdi2021critique}, it was proved how the tools and metrics that are employed to test contemporary model techniques and rank them in order of performance actually embody inconsistent and unreliable results. Some examples of the outcome: conflicting performance metric results, sporadic metric selection process and unsystematic model-data combinations for the experiments.

Building on those results, we focus in this new study on the key problem which is to understand and track the performance on small parts that makeup the whole dataset, and not just solely rely on general performance measures. The experiments in \cite{jurdi2021critique} show how randomly eliminating data from a dataset results in shockingly comparable performance results with algorithms that identify critical noisy data that have severe impact on performance. Even if not directly correlated, this aspect is in fact perfectly aligned with the research by Sun et al. \cite{sun2020we} where the authors recently uncovered some of the problems with evaluation and went on to prove that with the right setup and tools, a simple baseline model can outperform its superior and more complicated counterpart. While traditional evaluation mechanisms can provide a general indicator about how the performance is expected to be in a production environment, the overall model performance evaluation (such as a common train/split or cross-validation test) can fail to reflect the true behavior on smaller data groups of a dataset. This issue was presented in general domain-independent (non-recommender) machine learning studies such as the one conducted by McMahan et al. \cite{mcmahan2013ad} and recently thoroughly covered by Chung et al. \cite{chung2019automated} in their proposal of an automatic data slicing mechanism for the ultimate model evaluation process on the subset level \cite{chung2018slice,chung2019slice,chung2019automated}. The results show it is crucial to track the performance on a more granular level in order to really understand the effectiveness of a certain model based on the prediction objectives. Ultimately, this helps us better assess the effectiveness of a model on all parts of the available data for training and testing.

The concept of data slicing to uncover hidden severe performance fluctuations proved to be very effective \cite{chung2019automated}. This concept hasn't been utilized in the recommender system domain albeit being briefly touched on in the study about natural noise and the underlying performance fluctuations in some dataset parts \cite{jurdi2021critique}. In this study, we focus on adapting a tailored and modular (employing interchangeable core modules, more on that in Section \ref{sec:system_architecture}) slicing mechanism, called \textit{neighborhood-based evaluation}, to evaluate the performance of recommender models on a more granular level and based on the recommender's objective (ranking vs. prediction). The proposed mechanism evolved from the concept in \cite{chung2019automated} and works through clustering the dataset into neighborhoods and then using several techniques to identify the neighborhoods that a model performs the worst on. Neighborhood-based clustering in recommenders will allow the possibility to track model performance across dataset clusters and it can be utilized in several applications:
\begin{itemize}
    \item Enhancing automated decisions for model evolution.
    \item Noise/Fraud detection, such as malicious and natural noise.
    \item Creating a dynamic hybrid model structure in a production environment.
\end{itemize}

The main contribution of this paper is implementing a modular evaluation framework for the model validation process in recommender systems. The method is based on neighborhood clustering and analysis to identify weaknesses in model performance on certain groups of data with such weaknesses being usually hidden from metrics that report the overall performance. This research extends the previous work done on evaluation and noise management \cite{jurdi2021critique} and connects the evaluation slicing theory in \cite{chung2019automated,chung2019slice} to the recommender domain. 

The remainder of the article is arranged as follows: Section two introduces the neighborhood-based evaluation mechanism with a detailed description of the proposed approach which comprises a clustering technique, a neighborhood evaluation process and a metrics visualization method. Section 3 contains an applied investigation of the neighborhood-based mechanism that includes experimentation on several models, datasets, and evaluation metrics. Section 4 presents the possible application of the neighborhood-based evaluation approach and Section 5 provides a presentation on the the state-of-the-art works on the latest evaluation approaches in the field. Conclusions as well as a list of areas where we feel future work is particularly warranted are provided in Section 6.


\section{Neighborhood-based Evaluation - System Architecture}
\label{sec:system_architecture}
\begin{figure}[h]
  \caption{Neighborhood-based method architecture: Data slicing is performed to identify top-n problematic neighborhoods with metric performance visualizations}
  \centering
\includegraphics[width=0.8\textwidth]{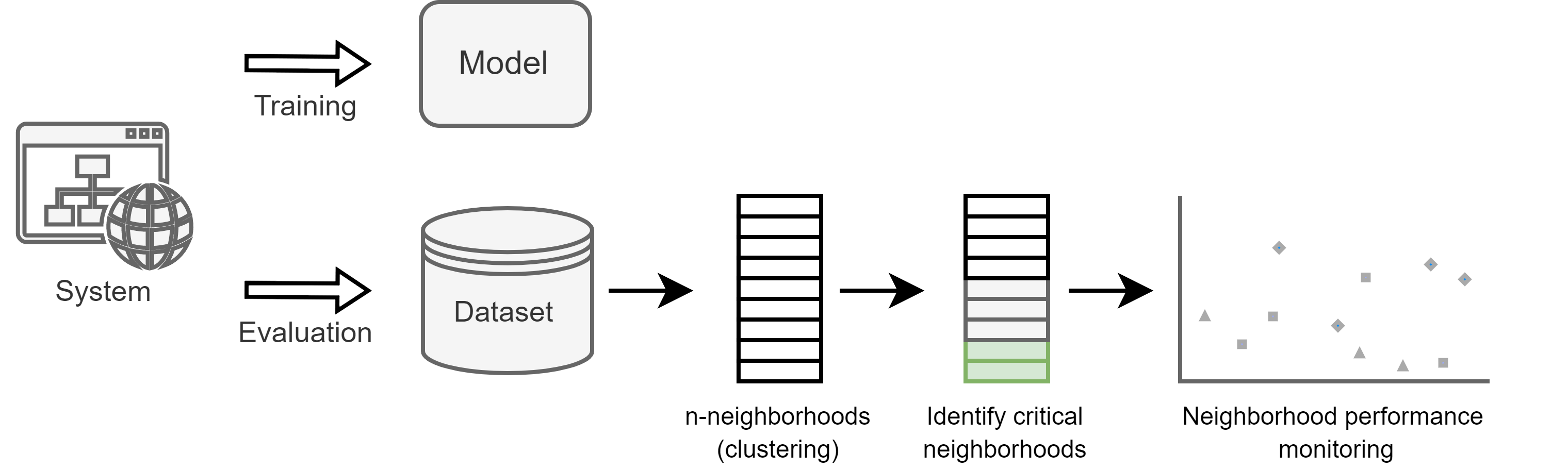}
  \label{fig:system_architecture}
\end{figure}

The neighborhood-based architecture, which is depicted in Figure \ref{fig:system_architecture}, is a modular framework that is inspired by a data evaluation slicing technique introduced in \cite{chung2019automated} and that works on the evaluation phase of a recommender training life cycle. Our approach is built specifically for recommender systems rating-based datasets and uses a different strategy (clustering-based) for forming groups compared to the one utilized in \cite{chung2019automated} to form data slices. The full details about the differences between the two methods as well as the automated slicing mode-of-operation presented by Chung et al. \cite{chung2019automated} are laid out in Section \ref{sec:sota_data_slicing}. Our source code\footnote{The public repository for the neighborhood-based evaluation mechanism as well as the experiment results output can be found here \cite{AlJurdi_nbhd_mechanism}. The experiments are in the form of Jupyter notebooks.} is Python-based and fully compatible with the models of Microsoft's new comprehensive algorithms repository for recommender systems \cite{argyriou2020microsoft}. This allows us to employ additional use-cases such as creating creating various model-dataset-metric combinations and utilizing/improving the method's core modules.

Consider a recommender model \(M\) and a dataset \(D\), a test metric \(h\) is a function that validates the performance of \(M\) on \(D\). Our main target in the neighborhood evaluation mechanism is to assess whether \(h\) is showing acceptable performance results for different slices \(N\), referred to as \textit{neighborhoods} in this context and represented in Figure \ref{fig:system_architecture}, of the whole dataset \(D\). Neighborhoods are formed through a chosen k-nearest neighbors algorithm (KNN-based) clustering technique. Our system searches for the neighborhoods that are in critical state and presents metric visualization plots which aid in tracking the overall performance results across all of those neighborhoods. The methodology followed for the neighborhood-based evaluation process can be summarized in the following steps:

\begin{enumerate}
    \item User-based clustering: A dataset is split into possibly overlapping neighborhoods using a KNN clustering method 
    \item Neighborhood evaluation:
    \begin{enumerate}
        \item Neighborhoods with a statistically significant higher loss than that of D are reported as critical.
        \item Using relevant evaluation metrics, the performance of the model on the neighborhoods is measured.
        \item Performance trends are visualized on the critical neighborhoods to identify and track the severity.
    \end{enumerate}
\end{enumerate}

\subsection{KNN Clustering process}
\label{sec:clustering_process}
In order to cluster users in the recommender datasets, we apply a KNN similarity-based clustering approach. Each resulting subset of the whole dataset will be considered as a stand-alone neighborhood and will be used in the testing and evaluation process which will be described in section \ref{sec:nbhd_evaluation_process} below. In order to decide which similarity measure to adopt as part of the grouping process of our neighborhood-based evaluation method, we conduct an initial experiment in section \ref{sec:experimentation} that shows the different results achieved with varying the grouping mathematical equation. The mean squared deviation measure (MSD), which will be referred to as KNN-1, will be tested along with other approaches: cosine (COS - KNN-2), pearson correlation coefficient (PCC - KNN-3) and pearson baseline correlation coefficient (PBC - KNN-4). PBC mainly uses baselines for centering instead of means; the theory behind this similarity distance measure can be reviewed in the \nth{5} chapter of \cite{rshandbook}. As will be shown in the first experiment, the system is actually capable of supporting any of the above methods. However, for conciseness, we adopt with the other experiments only one of the measures for the different group formation of the dataset in the neighborhood-based mechanism. All method formulations can be found in the following documentation \cite{surprise:similarity_measures}. The below formula represents PCC, the adopted mechanism within the neighborhood-based evaluation approach for all the experiments in this study:

\begin{equation}
    \label{eq:clustering_similarity}
      PCC(u,v) =
        \frac {\sum_{i\epsilon I_{u,v}} (r_{u,i}-\mu_u)(r_{v,i}-\mu_v)}
            {\sqrt{\sum_{i\epsilon I_{u,v}}(r_{u,i}-\mu_u)^2}\sqrt{\sum_{i\epsilon I_{u,v}}(r_{v,i}-\mu_v)^2}}
\end{equation}

Where \(r_{u,i}\) and \(r_{v,i}\) represent the ratings on item \(i\) by users \(u\) and \(v\) respectively. \(\mu_u\) and \(\mu_v\) are the average rating occurrences for both users, and \(I_{u,v}\) represents all the interaction items for these two users in the dataset. As mentioned in the introduction of section \ref{sec:system_architecture}, the clustering mechanism that's adopted in the neighborhood-based evaluation system is one of the core modules that can be altered and expanded to cover other methods such as the ones introduced in \cite{chung2019slice,chung2019automated,macdonald2021simpson} (more details on this can be found in section \ref{sec:a_modular_framework} that explains the modularity of our system). The main purpose of our study is to introduce a first version of the neighborhood-based evaluation framework in recommender systems. For brevity, we will follow the mentioned PCC clustering method for all the experiments as it works on all the rating-based datasets, in addition to a variety of recommendation algorithms that function on both implicit and explicit data. Other methods for other dataset types will be covered in a future extension of this work.

\subsection{Neighborhood evaluation process}
\label{sec:nbhd_evaluation_process}
The goal of the neighborhood evaluation is to identify and report data groups that exhibit bad performance and might be problematic to the system. As touched on in the introduction, it has been proven that such critical performance issues are usually hidden from the conventional evaluation results \cite{jurdi2021critique, macdonald2021simpson, chung2019slice}. The evaluation of neighborhoods is accomplished through the application of several tests that include a portfolio of different evaluation metrics that are selected based on the recommender's optimization, i.e., the recommender goal within an application's context \cite{ricci2021challenges}. Recommenders are normally of two main categories: optimized for effective predictions or optimized for effective ranking \cite{argyriou2020microsoft, surpriselib, aggarwal2016recommender}. Our evaluation method takes this into consideration in order to solve the evaluation issues/weaknesses presented in the introduction and avoid the conflicting and false results. These weakness we are focusing on are the ones discussed in the case studies presented in \cite{jurdi2021critique,macdonald2021simpson,sun2020we} which mention or include general pitfalls when evaluating models, inappropriate evaluation metric selection, and unclear view of the recommender's goal within a specific application.

To determine the performance status of every neighborhood \(N\) and identify the critical ones, we define a sub-dataset equivalent \(D'\) as \(D - N\) that encompasses the remaining data in a given dataset \(D\). For every data pair \( (N,D')\), we flag a neighborhood \(N\) as critical if \(N\) has a statistically significant higher loss than \(D'\). Statistical significance measures the existence of an effect whereby \(N\) indeed has a higher loss compared to its counterpart \(D'\) and the result is not just based on data variability and chance. To measure the statistical significance and identify the critical neighborhoods, we conduct several tests on every data pair \( (N,D')\) in a given dataset \(D\).

\subsubsection{Prediction/Classification loss}
\label{sec:prediction_loss}
In the first test towards identifying the critical neighborhoods, we assume a loss function \(\phi(N)\) that returns a performance score for a dataset by comparing a prediction \(r'_{u,i}\) with its true label \(r_{u,i}\), where \(r_{u,i}\) represents the rating of a user \(u\) on an item \(i\) in the recommender dataset. A common loss function will be utilized for the algorithms optimized for prediction according to the following equation:

\begin{equation}
    \label{eq:prediction_loss}
    \phi(N, D') = \sum_{r\epsilon (N,D')} (r_{u,i} - r'_{u,i})^2
\end{equation}

As for the algorithms optimized for the most effective ranking, we will use precision (\( Precision@k \)) that's defined as follows:

\begin{equation}
    \label{eq:ranking_loss}
    \phi(N, D') = \frac{TP}{TP + FP}
\end{equation}

In the recommender systems domain, \(TP\) would be the total recommended items that are relevant for a target user while \(TP + FP\) would represent the total number of recommended items. An item is considered relevant if its true rating \(r_ui\) is greater than a given threshold, whereas an item is considered recommended if its estimated rating \( \hat{r}_ui \) is greater than the threshold, and if it is among the k highest estimated ratings.

The system computes the relative loss of \(N\) and \(D\) according to equation \ref{eq:prediction_loss} or \ref{eq:ranking_loss} as the difference \(\phi(N) - \phi(D')\) and only considers the neighborhoods that result in a positive difference (Of course, the operation is switched between the two equations \ref{eq:prediction_loss} and \ref{eq:ranking_loss}). Naturally, this would imply that the loss in \(N\) is higher than that in \(D'\). Equations \ref{eq:prediction_loss} and \ref{eq:ranking_loss} could in fact be replaced by other functions (as will be touched on in section \ref{sec:a_modular_framework}), however, this level of process optimization to the clustering mechanism will be out of scope for this study and will be realized in a subsequent work.

\subsubsection{Statistical significance}
\label{sec:statistical_significance}
The second test will be conducted on the resulting neighborhoods of the prediction/classification loss test, and the goal is to measure the statistical significance of the difference between the loss of \(N\) relative to \(D'\) (for higher losses). For this purpose, we utilize the hypothesis t-testing with \(\phi(N) \le \phi(D')\) as the null hypothesis and  \(\phi(N) > \phi(D')\) as the alternative one. Since the two independent samples \(N\) and \(D'\) don't have identical average values, we perform the Welch’s version of the t-test \cite{enwiki:1087703061} which does not assume equal population variance. This will make our case similar to that in \cite{chung2019automated} where the data slices also do not have identical average values. The data pairs that pass this test after the first will be considered in the critical zone and will be part of the evaluation and monitoring process.

\subsubsection{Evaluation metrics}
In this third test, which is mostly a visualization procedure, we track the performance on the neighborhoods that are flagged as critical based on the results of the above tests. To accomplish this, the final list of critical neighborhoods will be passed into an evaluation process using several metrics that are usually employed for the overall system evaluation. The chosen metric profiles will vary depending on the goal of the system and what we are evaluating for, and they will include: accuracy measures for models optimized for prediction tasks, and ranking-based measures for algorithms optimized for effective rankings \cite{tamm2021quality,aggarwal2016recommender}.

\begin{definition}
\label{def:critical_neighborhood}
Critical neighborhood: A group of data points reported by the proposed neighborhood-based evaluation mechanism. This data portion exhibits significantly worse results, through a certain recommender algorithm and as measured by certain metrics, compared to the other examples in the dataset. 
\end{definition}

\section{Experimentation}
\label{sec:experimentation}

\subsection{Experimental setup}
In this section we present the findings and the overall experimental processes/procedures and findings of the neighborhood-based evaluation mechanism. The experiments are split into three main categories where in the first one, we introduce a test where the clustering algorithm/parameter of the designed system is varied with the goal of showcasing different grouping options and the performance change with each. In the second and third experiments, we showcase the implementation of the neighborhood-based evaluation using several recommendation algorithms and different datasets.

\subsubsection{Datasets and Algorithms}
The datasets utilized in this study are mainly rating-based: ml-latest-small, ml-latest, ml-1m \cite{harper2015movielens} and personality \cite{nguyen2018user}. Table \ref{tab:datasets} summarizes the information about the datasets including their sparsity. In the examples where ml-latest was used, we opted for a smaller version of it referred to as ml-latest*.

\begin{table}[h]
  \caption{Datasets used in the experiments of the Neighborhood-based evaluation mechanism.}
  \label{tab:datasets}
  \begin{tabular}{ccccl}
    \toprule
    Dataset & Total users & Total items & Total ratings & Sparsity \\
    \midrule
    ml-latest-small & 610 & 9,742 & 100,836 & 0.983 \\
    ml-1m & 6,040 & 3,900 & 1,000,209 & 0.957 \\
    personality & 1,820 & 35,196 & 1,028,751 & 0.983 \\
    ml-latest & 283,228 & 58,098 & 27,753,444 & 0.998 \\
    ml-latest* & 16,000 & 24,316 & 1,571,685 & 0.995 \\
  \bottomrule
\end{tabular}
\end{table}

Recommender algorithms could be categorized into two broad classes: those optimized for accurate predictions and others optimized for the finest rankings. For each of these two categories, we focus on measuring what better reflects the goal of the recommender. In the case algorithms are optimized in a certain application for accurate predictions, we utilize accuracy evaluation instead of a combination between accuracy and ranking tests. The neighborhoods can adjust to any of the needed tests but we will be applying the test that better fits the recommender's goal or category for each algorithm (recall Section \ref{sec:system_architecture}).

\subsubsection{Algorithms designed for effective predictions}
Currently becoming less popular in online settings, especially e-commerce \cite{jurdi2021critique}, such algorithms are optimized to have the most accurate prediction of a rating a user could provide to a new item.
\begin{itemize}
    \item SVD \cite{svd:simon_funk, surpriselib}. A matrix factorization algorithm. When baselines are not used, this is equivalent to Probabilistic Matrix Factorization. The prediction is defined as: \( \hat{r_{u,i}} = \mu + b_u + b_i + q_{i}^{T}p_u \). To estimate the unknown ratings, a regularized squared error is minimized using a straightforward stochastic gradient descent.

    \item SlopeOne \cite{lemire2005slope}. A simple yet very accurate collaborative filtering algorithm where the prediction is defined as: \( \hat{r_{u,i}} = \mu_u \frac{1}{|R_i(u)|} \sum_{j \epsilon R_i(u)} dev(i,j) \). The \( dev(i,j) \) is the average difference between the ratings of \(i\) and those of \(j\).

    \item NMF \cite{luo2014efficient}. A collaborative filtering algorithm based on Non-negative Matrix Factorization and that is very similar to SVD \cite{svd:simon_funk}. The prediction is defined as: \( \hat{r_{u,i}} = q_{i}^{T} p_u \). The optimization procedure is a regularized stochastic gradient descent with a specific choice of step size that ensures non-negativity of factors, provided that their initial values are also positive.
\end{itemize}

\subsubsection{Algorithms designed for effective rankings}
In contrast to prediction-optimized algorithms, personalized ranking aims at providing each user a ranked list of recommender items. It's a common scenario when the system's user behavior is based on implicit data. Although the focus of this study is mainly the rating-based datasets in the neighborhood evaluation approach, we will experiment on algorithms designed for effective rankings to provide a complete initial framework. Therefore, in our case, the system's clustering approach for neighborhood formation is only relevant to the rating-based dataset type.

\begin{itemize}
    \item SVDpp \cite{surpriselib, niu2018collaborative}. A matrix factorization algorithm that is extended from SVD \cite{svd:simon_funk} while taking into account implicit ratings. The prediction is defined as: \( \hat{r_{u,i}} = \mu + b_u + b_i + q_{i}^{T} (p_u + |I_u|^{-\frac{1}{2}} \sum_{j \epsilon I_u} y_j) \). This additional term in the equation refers to the fact that when a user rates an item, it will be an indication of preference. In other words, chances that the users "like" an item they have rated are higher than for a random not-rated item.

    \item BPR \cite{rendle2012bpr}. This model uses item pairs \( i,j \) and optimizes for the correct ranking given the preference of a user \(u\) by maximizing the posterior probability. The final objective function of the maximum posterior estimator is: \( J = \sum{(u,i,j)\epsilon D_s} ln \sigma(\hat{x}_{uij}) - \lambda_\theta ||\theta||^2 \).
\end{itemize}

\begin{figure}[h]
  \caption{Percentage of reported critical neighborhoods with varying the KNN clustering method on different datasets.}
  \centering
\includegraphics[width=0.55\textwidth]{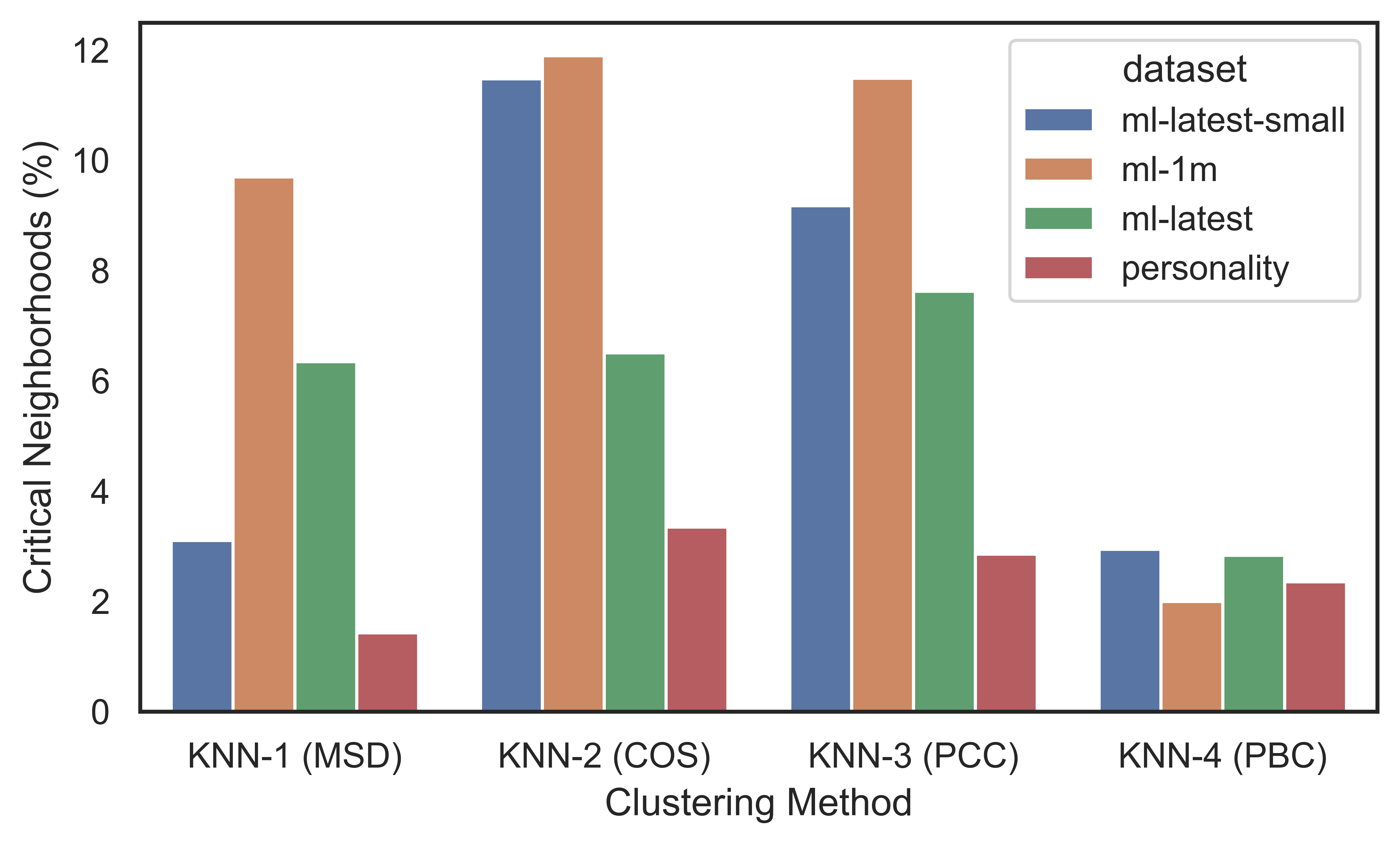}
  \label{fig:all_knn_critical_nbhds}
\end{figure}

\begin{figure}[h]
  \caption{MSE, MAE and RMSE (prediction error metrics) results for dataset pairs \( (N,D')\) using different KNN clustering methods (left to right) MSD, COS and PCC. Dataset: ml-latest-small.}
  \centering
\includegraphics[width=0.85\textwidth]{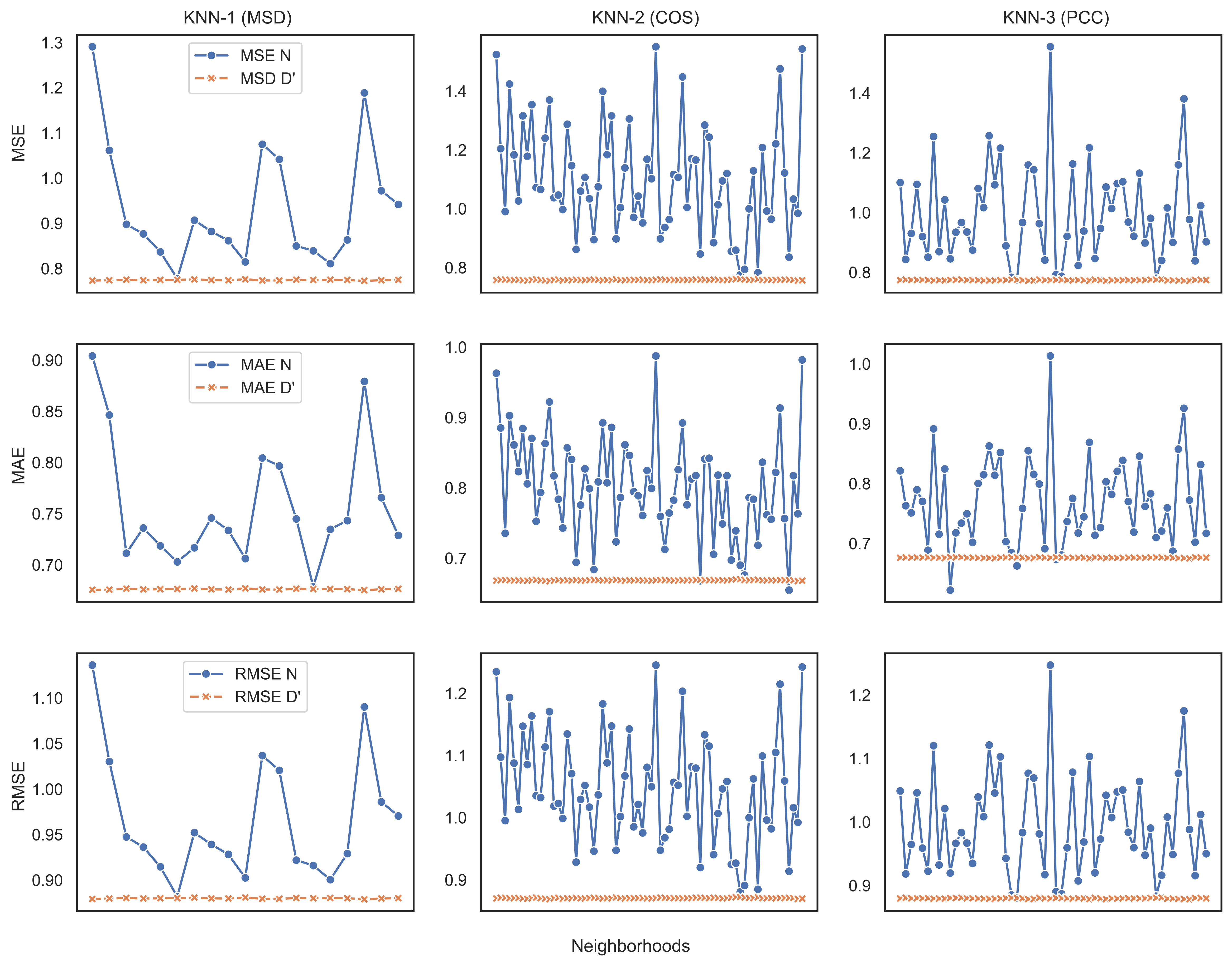}
  \label{fig:knn_methods_ml_latest_small}
\end{figure}

\subsection{KNN Clustering methods}
\label{sec:knn_clustering_methods}
In this initial supplementary experiment, we apply the similarity-based KNN clustering methods that were described in section \ref{sec:clustering_process} then compare them to each other. We used multiple datasets: ml-latest-small, ml-1m, ml-latest and personality. With respect to the recommender algorithm, we fixed an SVD model coupled with the accuracy metrics MSE, MAE and RMSE. The main goal of this experiment is to provide an overview on the neighborhood evaluation framework and its \textit{modularity} (recall Section \ref{sec:system_architecture}), as well as present the variations that could result from altering the nearest-neighbor method in the KNN clustering approach for neighborhood formation. In our full offline experiments, and in a more expanded approach, we tested other datasets with the KNN methods \cite{AlJurdi_nbhd_mechanism}. The results were almost consistent with each other and in this work we will report the most important findings while keeping a copy of all the outcomes in the source code \cite{AlJurdi_nbhd_mechanism}.

In its first phase, the neighborhood evaluation mechanism calculates the prediction loss using Equation \ref{eq:prediction_loss}. After that, the second test proceeds to measure the statistical significance (based on Welch's t-test variant - recall Section \ref{sec:statistical_significance}) of the difference between each neighborhood's loss and that of the rest of the dataset but only for neighborhoods with higher losses (i.e. for positive differences). Figures \ref{fig:all_knn_critical_nbhds} and \ref{fig:knn_methods_ml_latest_small} summarize the results of this experiment, where the first reports the percentage of critical neighborhoods out of the total neighborhoods (on multiple datasets) that were formed based on the selected KNN clustering method, while the second shows a scatter plot of the critical neighborhoods with their metric values. Analyzing the results of Figure \ref{fig:all_knn_critical_nbhds}, the lowest reported number of critical neighborhoods is when KNN-4 (PBC) is utilized which is normal as the measure is using baselines for centering instead of means and that's affected by the sparsity of the datasets. The other three methods are showing relatively similar results where KNN-1 (MSD) is on the lower end of the score with around 8\% of reported critical neighborhoods on the ml-latest-small dataset and 13\% for both the ml-1m and the ml-latest datasets. KNN-3 (PCC) is slightly on the higher end of the results, however, displaying more consistent results between the four datasets compared to KNN-2 (COS). KNN-2 generally resulted in higher number of reported critical neighborhoods in our experimentation with around 15\% and 16\% of neighborhoods reported as critical for ml-latest-small and ml-1m respectively. 

The scatter plots of Figure \ref{fig:knn_methods_ml_latest_small} display the metric results (MSE, MAE, and RMSE) for the three KNN clustering techniques for each \((N, D')\) pair, where each point corresponds to one of the critical neighborhoods. It's evident how the critical neighborhoods have higher losses or error than their counterparts \(D'\) (most showing much higher values than \(D'\)) with these errors being in the critical zone i.e. above 0.85 for MSE, above 0.75 for MAE, and above 0.90 for RMSE. \(D'\) has a consistent score across all the neighborhoods that is close to the system's metric result (The MSE, MAE and RMSE that are evaluated on the whole data). This is an indicator that the accuracy of predictions of the SVD algorithm on the critical neighborhoods shown in Figure \ref{fig:knn_methods_ml_latest_small} is low compared to the prediction performance on the rest of the user groups of the dataset. Since the metrics almost measure the same goal in this case, we adopt RMSE for the general performance of the other experiments as we vary the recommender algorithms with the neighborhood evaluation mechanism.

As previously touched on in the method introduction in Section \ref{sec:system_architecture}, the aim of this work is to introduce the neighborhood-based evaluation framework and propose an evaluation mechanism on a more granular level of the data, taking into account the goals of the recommender. The distance measure for neighborhood formation is an essential process that's modifiable and will be investigated further in future work. For the following experiments, we fix KNN-3 (PCC, Equation \ref{eq:clustering_similarity}) as the core method for KNN clustering mechanism.

\begin{figure}[h]
  \caption{Percentage of critical neighborhoods reported with every prediction-based recommender on different datasets.}
  \centering
\includegraphics[width=0.55\textwidth]{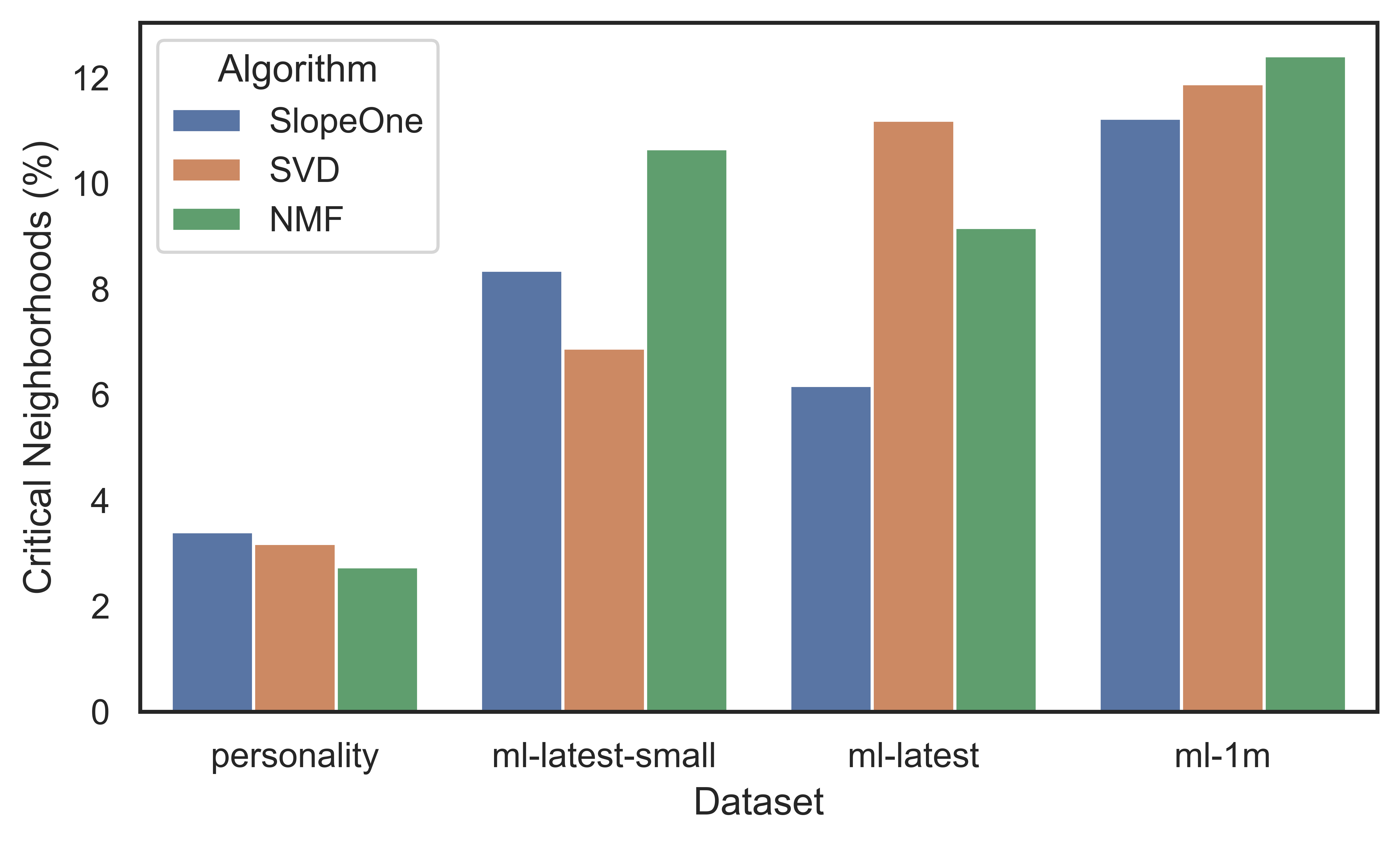}
  \label{fig:exp_2_critical_nbhds}
\end{figure}

\subsection{Prediction algorithms}
In the second experiment, we target implementing the neighborhood evaluation process on algorithms designed for optimal prediction. The selected models are SVD, SlopeOne and NMF. The goal of this experiment is to showcase the change in the percentage of reported critical neighborhoods (recall Definition \ref{def:critical_neighborhood}) as the algorithm is varied, in addition to whether some algorithms could potentially perform better than the others on the neighborhoods flagged as critical by them. What's also interesting to investigate is the common critical neighborhoods between algorithms when using the neighborhood-based evaluation process.

\begin{figure}[h]
  \caption{Scatter plot for the RMSE results of the dataset pairs \( (N,D') \) with different recommendation algorithms (SVD, SlopeOne and NMF) using two datasets (ml-latest-small and personality}
  \centering
\includegraphics[width=0.8\textwidth]{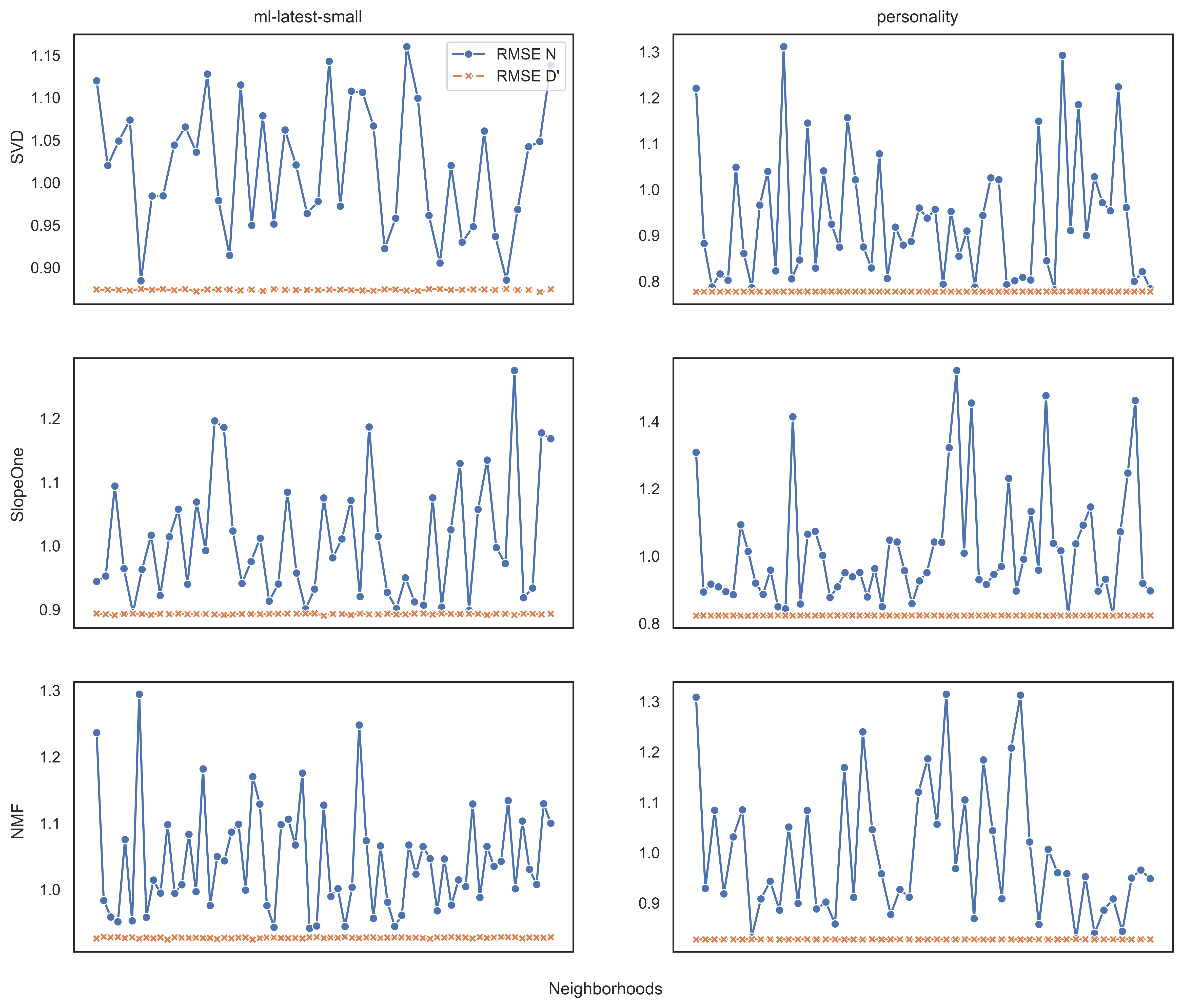}
  \label{fig:exp_2_scatter}
\end{figure}

\begin{figure}[h]
  \caption{Box plot of the dataset pairs \( (N,D') \) based on the RMSE values using three different recommender algorithms and two datasets (ml-latest and ml-1m).}
  \centering
\includegraphics[width=0.6\textwidth]{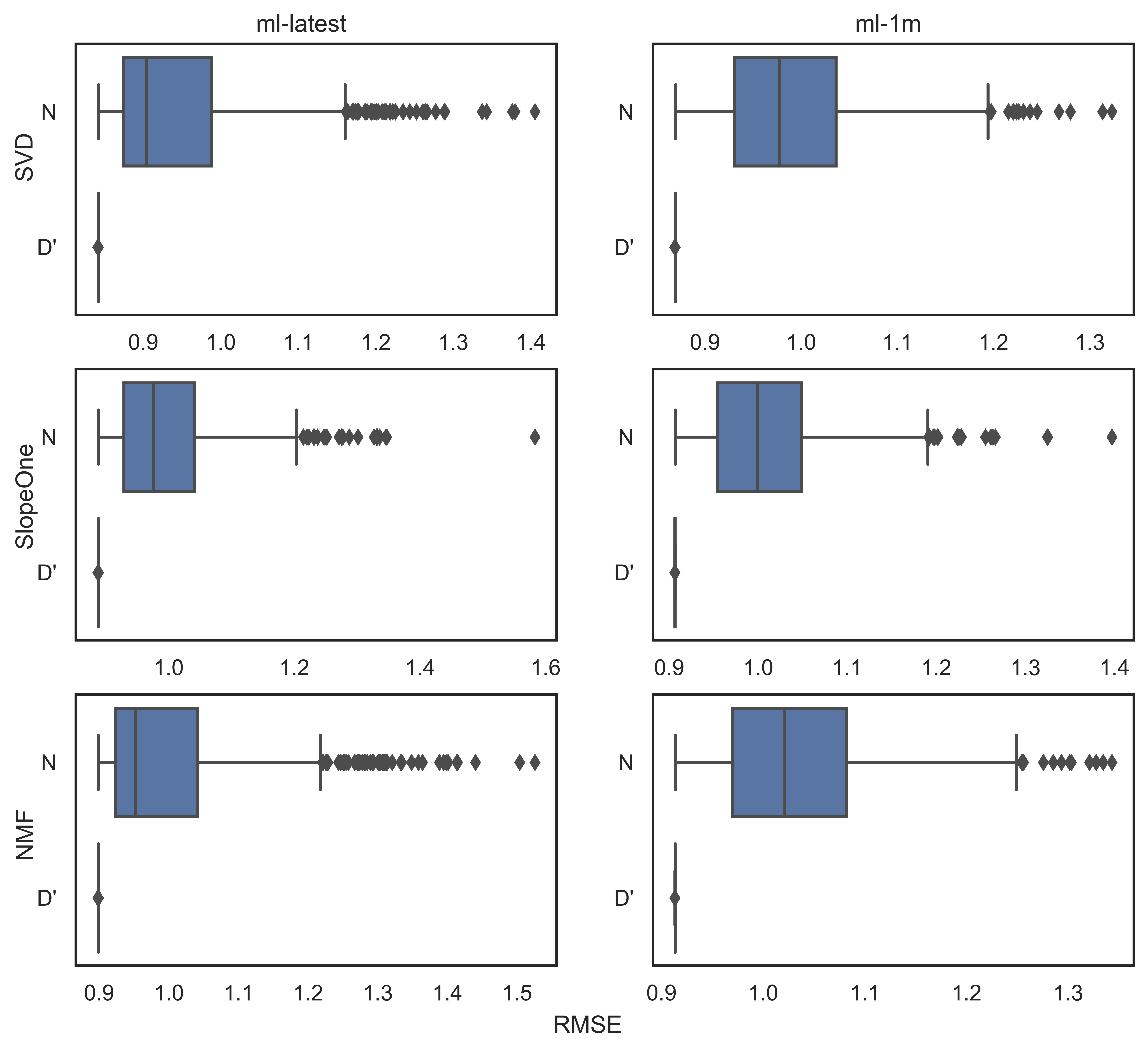}
  \label{fig:exp_2_box}
\end{figure}

The variations of the percentage of critical neighborhoods with every recommender in shown in the bar graph of Figure \ref{fig:exp_2_critical_nbhds}. Personality still scores the lowest and the most stable with all the algorithms (around 3.8\%), and that's due to the nature of the dataset. It was optimized for accurate user personalities and contains reliable data which in this case would reduce the number of critical reported data groups (low loss of \(N\) compared to \(D'\)). The scatter plot in Figure \ref{fig:exp_2_scatter} shows the reported critical neighborhoods on two datasets, ml-latest-small and personality. Since the total number of critical neighborhoods for those two datasets is relatively small, the representation in this figure gives us the ability to better illustrate  the RMSE variations that those neighborhoods experience compared to the equivalent data points \(D'\) . It is clear from the figure that \(D'\) RMSE values are almost constant in all the neighborhood cases while the critical neighborhoods \( N \) vary on the higher RMSE level. Some of the neighborhoods scored error values as high as 1.6 in many cases for both datasets and with all the three recommender algorithms. In a clearer representation of the high error values that the critical neighborhoods are experiencing, Figure \ref{fig:exp_2_box} illustrates in a box plot the critical neighborhoods' RMSE values for the other two datasets using the same recommender algorithms. Comparing the boxplot of \(N'\) to \(D\) for each recommender algorithm and dataset, we can clearly see that the range of RMSE values (Q1, median, Q3, and Interquartile range - IQR) is much higher (the box is shifted remarkably to the right) for \(N'\) confirming that the difference in RMSE between the critical neighborhoods and their \(D'\) equivalent is significant with higher values for the neighborhoods. We can also notice on the figure many neighborhoods (data points) as outliers to the right of the maximum of the \(N\) box indicating their much higher RMSE values even compared to their neighbors' distribution. The narrowness of the box for \(D'\) in all figures confirms the tight range of RMSE values for \(D'\) i.e. \(D'\) exhibiting very close RMSE values for all neighborhoods and cases.

In order to investigate the common critical neighborhoods between the different recommender algorithms in the experiment, we record the percentage of critical neighborhoods that are common between the three algorithms, between any two of them, and the ones that are unique to just one. Table \ref{tab:exp_2_critical_nbhds} shows the values registered for this test. We can notice from the table that the percentage of common critical neighborhoods between the three algorithms (SVD, SlopeOne and NMF) and any two algorithms is relatively very low. This implies that different algorithms are performing differently on the neighborhood clusters. The highest percentages (between \(88\%\) and \(99.5\%\)) are recorded for the unique critical neighborhoods in each algorithm alone. This shows that the recommendations on the neighborhoods have varied performance results where a neighborhood that has high error within one model tends to be in a better position when trained using another. This open up the possibility of using a hybrid model system where we could have several models catering for different neighborhood groups (more on that in \ref{sec:hybrid_model_decision_system}.

\begin{table}[h]
\centering
\caption{Distribution of critical neighborhoods between the three algorithms (SVD, SlopeOne, and NMF) for the three datasets.}
\label{tab:exp_2_critical_nbhds}
\begin{tabular}{llll} 
\toprule
\multirow{2}{*}{ Datasets} & \multicolumn{2}{l}{Sum of common critical neighborhoods } & \multirow{2}{*}{\begin{tabular}[c]{@{}l@{}}Sum of unique critical neighborhoods\\in each algorithm alone\end{tabular}}  \\ 
\cmidrule{2-3}
                           & Between the 3 algorithms & Between 2 algorithms           &                                                                                                                         \\ 
\midrule
ml-latest-small            & 3.8\%                    & 13.92\%                        & 88.79\%                                                                                                                 \\
ml-1m                      & 2.94\%                   & 16.47\%                        & 88.58\%                                                                                                                 \\
ml-latest                  & 4.1\%                    & 19.98\%                        & 98.88\%                                                                                                                 \\
personality                & 2.03\%                   & 23.55\%                        & 99.42\%                                                                                                                 \\
\bottomrule
\end{tabular}
\end{table}

\subsection{Ranking algorithms}
\begin{figure}[h]
  \caption{Percentage of critical neighborhoods reported with the ranking-based recommenders on different datasets.}
  \centering
\includegraphics[width=0.55\textwidth]{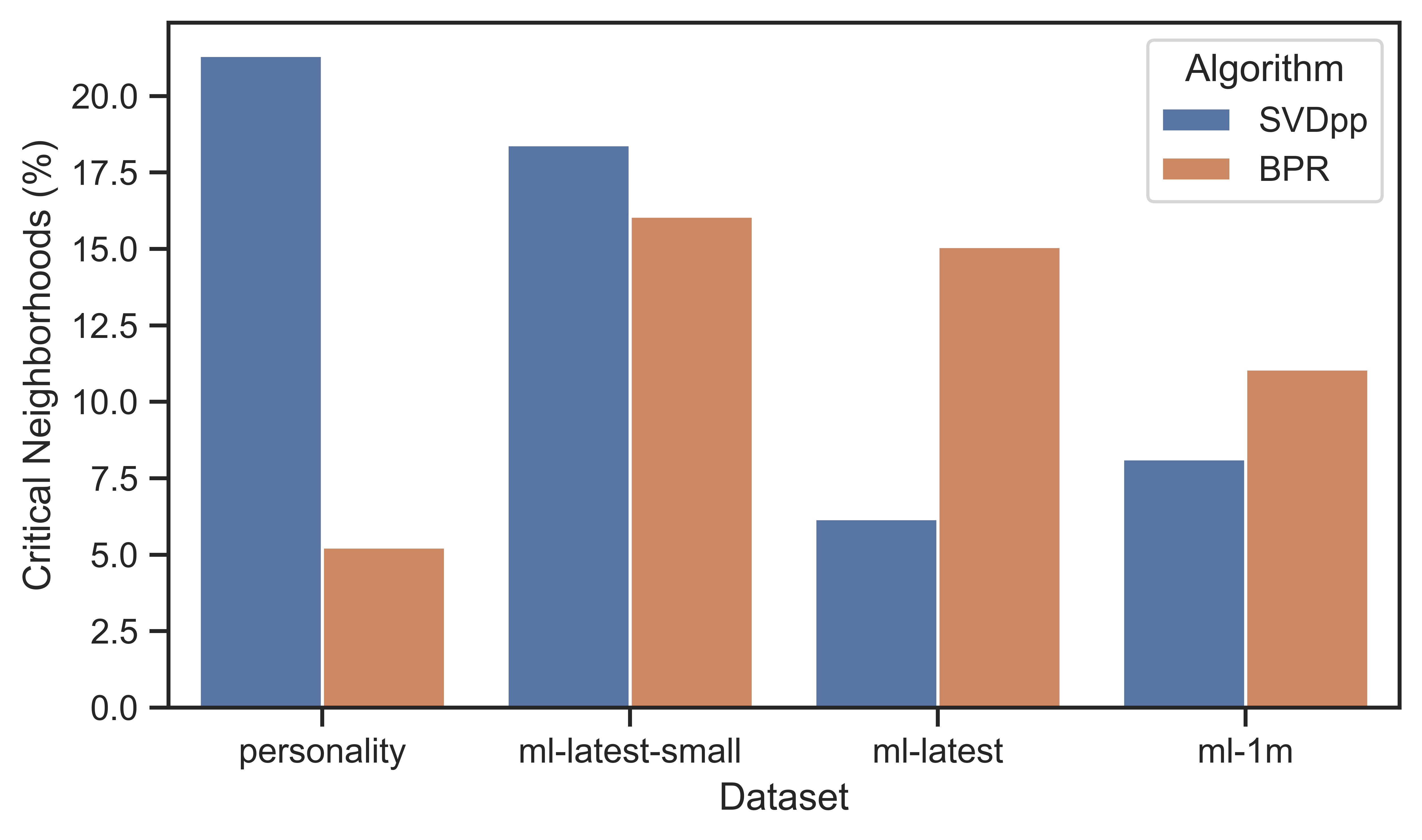}
  \label{fig:exp_3_critical_nbhds}
\end{figure}

The same strategy for the neighborhood formation and loss/statistical test is applied for this third experiment. In this case, we target implementing the neighborhood evaluation process on algorithms designed for optimal ranking. The selected models are SVDpp, BPR. The difference in this test is in the loss function used (Equation \ref{eq:ranking_loss}) and in the metrics we focus on to check the reported critical neighborhoods' performance. The variations of the percentage of critical neighborhoods with the two recommender models is shown in the bar graph of Figure \ref{fig:exp_3_critical_nbhds}. Interestingly, the highest number of report critical neighborhoods was in the case of the personality with the SVDpp algorithm (20\%). It seems that SVDpp doesn't perform that well with the optimized data in personality that includes a small amount of users compared a relatively high amount of items and ratings. The other critical neighborhood percentages are a bit higher compared to our previous test, which indicates that there are groups that are more affected with bad ranking results compared to prediction results.

\begin{figure}[h]
  \caption{Scatter plot for the F1 results of the dataset pairs \( (N,D') \) with different recommendation algorithms (BPR, F1) using two datasets (ml-latest-small and personality}
  \centering
\includegraphics[width=0.7\textwidth]{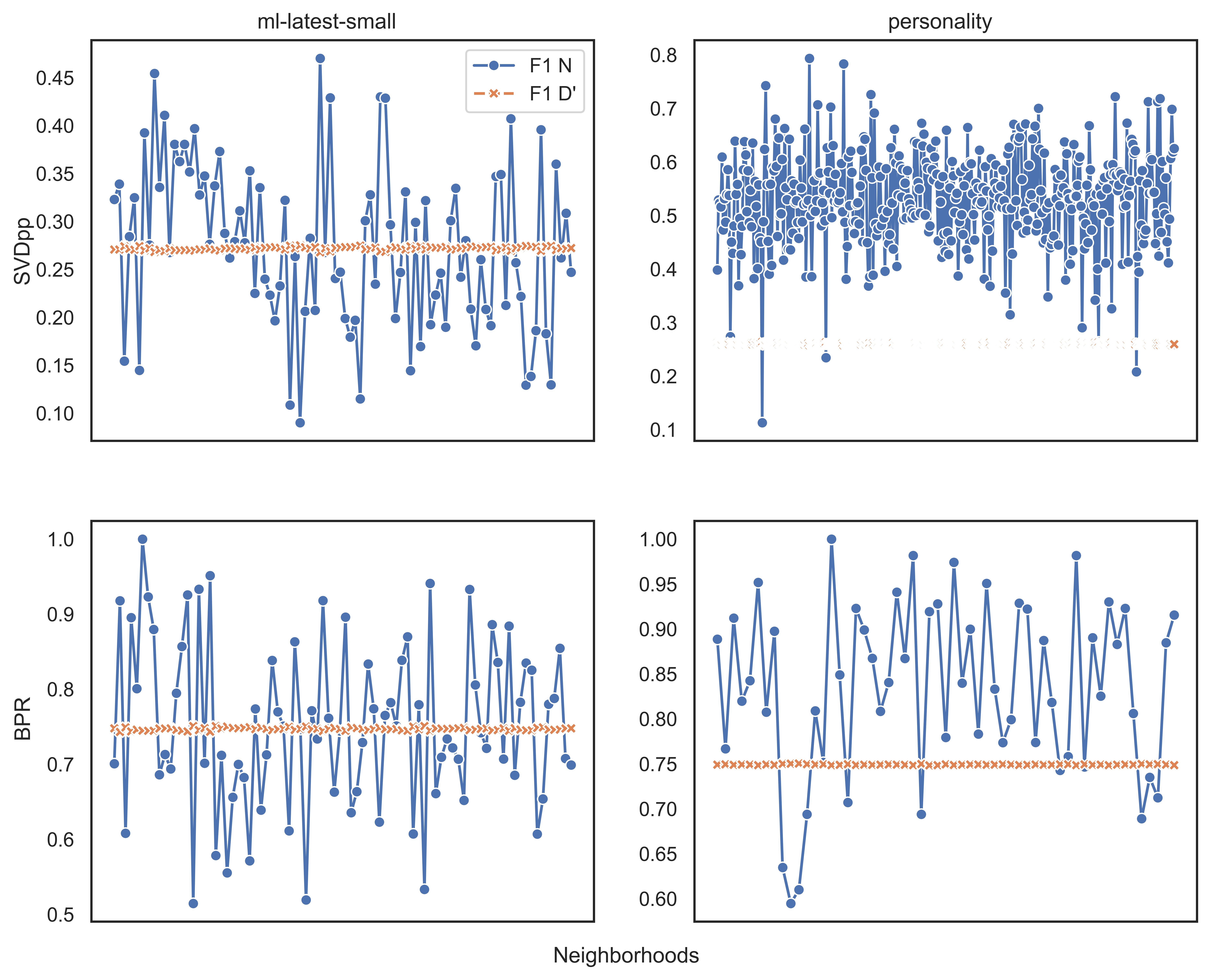}
  \label{fig:exp_3_scatter}
\end{figure}

The scatter plot in Figure \ref{fig:exp_3_scatter} shows the reported critical neighborhoods on two datasets, ml-latest-small and personality. Since the total number of critical neighborhoods for those two datasets is relatively small we also keep the same representation as the one we showed for the prediction-based algorithm in Experiment 2. It is clear from this plot that \(D'\) F1 values are also almost constant in all the neighborhood cases while the critical neighborhoods \( N \) still greatly vary, but this time in a different way compared to the previous experiment.

Some of the neighborhoods scored error values as high as 1.6 in many cases for both datasets and with all the three recommender algorithms.

\begin{figure}[h]
  \caption{Box plot of the dataset pairs \( (N,D') \) based on the RMSE values using three different recommender algorithms and two datasets (ml-latest and ml-1m).}
  \centering
\includegraphics[width=0.6\textwidth]{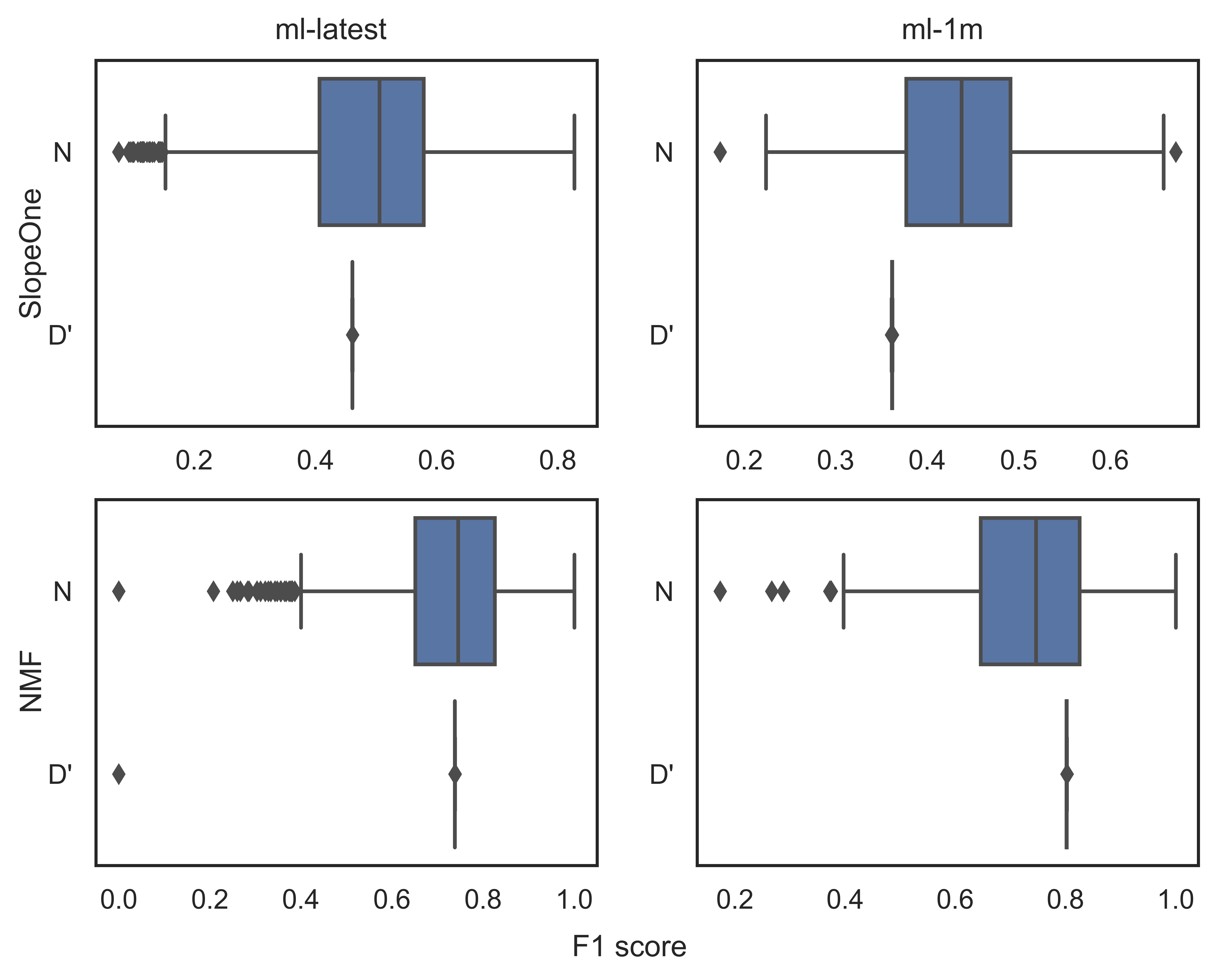}
  \label{fig:exp_3_box}
\end{figure}

\section{Neighborhood-based Evaluation Mechanism Applications}
\subsection{A modular framework}
\label{sec:a_modular_framework}
In this study, we introduced the neighborhood-based evaluation mechanism to the recommender systems ecosystem based on the similarity grouping (KNN). The methods used and implemented in our open-source repository \cite{AlJurdi_nbhd_mechanism}, like the similarity measure for user grouping, are easily interchangeable which makes the system modular. The goal is to expand the functionality of the system and make it compatible with all dataset types in the recommender domain.

\subsection{Model evolution and fairness}
With the recent vital importance of fair recommender systems \cite{singh2018fairness}, it is crucial to report and analyze the performance for a slice of users or items.

\subsection{Noise/Fraud detection}
Fraud/Noise detection involves identifying patterns of activities where a model is not performing as well as it previously did, which is possible with the neighborhood-based evaluation method for recommenders. For example, some natural noise patterns can be affecting parts of the system as touched on in \cite{jurdi2021critique}. Evaluating the performance on a more granular level will help spot those weaknesses and avoid false system training/recommendations.

\subsection{Hybrid Model decision system}
\label{sec:hybrid_model_decision_system}
\begin{figure}[h]
  \caption{Hybrid model proposal based on the neighborhood-based evaluation method.}
  \centering
\includegraphics[width=0.55\textwidth]{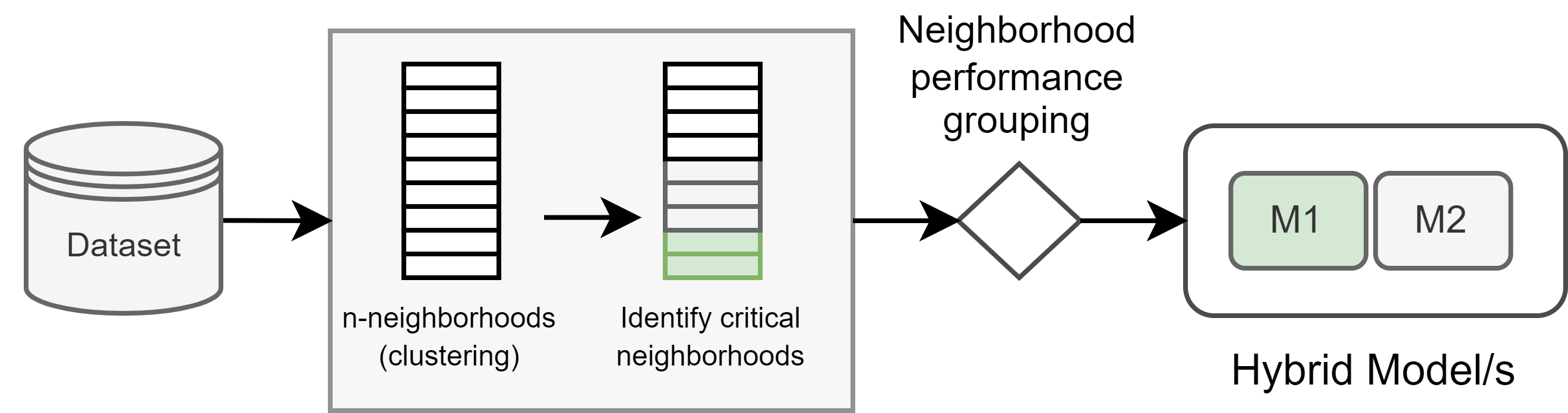}
  \label{fig:hybrid_models}
\end{figure}

Utilizing the neighborhood evaluation method introduced in this research, the system architecture could be further expanded to include multiple models deployed in a production environment where each would be tuned for different groups based on the results from the neighborhood-based system. Figure \ref{fig:hybrid_models} shows a flow diagram of this proposed application method. The critical neighborhoods could be fed into a grouping mechanism that arranges them based on the metrics portfolio similar to that employed in the experiments of Section \ref{sec:experimentation}. This is a useful application which aids in limiting degrading performance results on small groups.

\section{Background and Related Work}
\subsection{Data slicing - clustering vs automated processing}
\label{sec:sota_data_slicing}
A series of studies that focus on data slicing while also covering clustering methods for model evaluation was introduced by Chung et al. \cite{chung2018slice,chung2019slice,chung2019automated}. This series of work is the closest to the neighborhood-based method proposed in this research as both are leveraging the concept of evaluating performance on smaller data pieces as opposed to a full dataset evaluation. However, there are several core differences as we will describe in what follows. The authors' final method which was presented in \cite{chung2019automated} and called \textit{SliceFinder} is not developed in a way that would work for the recommender systems domain but rather for general classification problems. The reason for that is the use of the mechanism of features in the dataset to create subsets that make sense to the user in the end; such features are not present in recommender and personalization datasets where we usually have limited information about the users in an interactive system. SliceFinder is an interactive framework for identifying problematic slices using statistical techniques, and the slice evaluation mechanism is based on a general classification loss function defined as follows:

\begin{equation}
\label{eq:log_loss_data_slicing}
    - \frac{1}{n} \sum_{(x_{F}^{i} ,y^i) \epsilon S} [y^i ln  h(x_{F}^{i})+(1 - y^i) ln (x_{F}^{i}))]
\end{equation}

This classification loss function returns a performance score for a set of examples by comparing h’s prediction \( h(x_{F}^{i}) \) with the true label \(y^i\). The difference between the tests conducted to identify the least performing slices in this case and the method used for identifying the critical neighborhoods with our mechanism is the core loss function, as well as the evaluation metrics that were employed based on the recommender optimization and goal. In the neighborhood-based evaluation case, the loss function is connected to the evaluation metric that will eventually be used to measure the system's performance. Moreover, we define several loss functions that are based on the recommender system's optimization. As for the slicing mechanism, the authors in \cite{chung2019slice} applied several methods to implement an automated data slicing technique such as decision trees (non-overlapping) and lattice searching (overlapping) which also generated meaningful data chunks (also referred to as literals). Our neighborhood-based method is different in this regard as the dataset for recommender system are different and include unique features and normally much less information compared to the plenty of features utilized in the study of SliceFinder. We are currently using a neighborhood approached based on the minimal features that are available in the type of datasets used in the experiments of this study, and if we applied SliceFinder, we will end up with a lot of data points groups that don't make sense to the user (example of a slice that's reported by SliceFinder and makes sense: \(country = DE \cap gender = Male\)). Such features are usually lacking when dealing with recommender system datasets, however, in a future study, we will tackle this issue by generalizing the neighborhood-based architecture to cover implicit feedback data that's found in the majority of e-commerce applications.

In the proposal of a comprehensive and rigorous framework for reproducible recommender evaluation by Anelli et al. \cite{anelli2021elliot}, the authors mentioned the idea of statistical tests on data groups in the third section of the study. Throughout the work, it was emphasized that there's a need to compute fine-grained (such as per-user or per-partition) results and retain them for each recommendation model. As a result, their framework, called Elliot, was designed for multi-recommender evaluation and handling the fine-grained results. Elliot brings the opportunity to compute two statistical hypothesis tests, i.e., Wilcoxon and Paired t-test, activating a flag in the configuration file. The proposal didn't mention the partitioning techniques used, other than the idea that the partitions might be per user. Partitioning per user is a possibility in our proposed method and the neighborhoods formed by the methods are all centered around the user in the KNN clustering scheme. However, the main concept of neighborhood evaluation is to report partitions that make sense, i.e. neighborhoods. That way, the evaluation process gives the ability to check the worst performing groups and analyze the reason behind the results.

\subsection{The Simpson's paradox}
While not directly related to the evaluation of slices, the Simpson's paradox in the offline evaluation of recommendation systems (Jadidinejad et al. \cite{macdonald2021simpson}) introduces a phenomenon that describes the effects shown in the results of our study. The research in \cite{macdonald2021simpson} shows that the typical offline evaluation of recommender systems suffers from the phenomenon termed \textit{Simpson’s paradox}. Simpson’s paradox is when a significant trend appears in several different sub-populations of observational data but disappears or is even reversed when these sub-populations are combined together. This definition explains the results obtained in our experiments where the neighborhood-based evaluation mechanism was able to spot and report smaller versions of the data (neighborhood) where the performance of the model suffers. This negative performance will not be apparent when we evaluate the system with legacy evaluation methods.

Although the definition of the Simpson's paradox applies in our case, the authors in \cite{macdonald2021simpson} conducted different experiments and proposed an approach that tackles a different issue with the datasets of recommender systems. The experiments are based on "stratified sampling" and reveal that a very small minority of items that are frequently exposed by a deployed system (such as the system that helped generate the datasets in Table \ref{tab:datasets} that are used in most of the studies on recommender systems) plays a confounding factor in the offline evaluation of recommendation systems. So, the study investigates the issue of an initial recommender system (called the confounder) that influences the rating elicitation process. That's the different between the study in the Simpson's paradox and the neighborhood-based evaluation mechanism, where the latter only tackles the issue with the performance of the reocmmender on smaller groups of users.

\subsection{Benchmarking for evaluation}
A missing idea in the benchmarking proposals is the differentiation between different recommender algorithms and their general goal inside a certain application. In our proposal, the neighborhood-based mechanisms evaluates a recommender based on the optimization of the model, which is the recommended way of approaching performance assessment \cite{argyriou2020microsoft}. As touched on in the introduction of our work, with the increase in number of recommender algorithms proposed as well as different approaches to enhance them, a critical issue presents itself. There isn't a standardized approach to evaluate the algorithmic enhancements as recommender proposals evolve. Some of the recent studies on evaluation focus on the benchmarking approach in an attempt to create a framework for the most accurate evaluation and comparison. In one of the proposals by Sun et al. \cite{sun2020we}, the authors aimed to conduct rigorous (i.e., reproducible and fair) evaluation for implicit-feedback based top-N recommendation algorithms. They reviewed several recent proposals and analyzed the different approaches used for evaluating recommender systems. As expected, there were several inconsistencies in what was utilized for evaluation which led to inconsistent results when judging if a new proposal is better than its predecessor. Accordingly, the authors created benchmarks with standardized procedures and provided the performance of seven well-tuned models across six metrics on six widely-used datasets.

In a similar, and more recent study, Ovaisi et al. \cite{ovaisi2022rgrecsys} proposed a toolkit for evaluation that's targeted for assessing the robustness of recommender models. In this research the authors mentioned the necessity for evaluating the model on different slices of data such as gender subgroups. In one particular example of the study, a sub-population of the test set was formed and that consisted of users who identify as females and males. It was shown that the system performed much worse for females compared to males across all the models used in the experiment. This is another proof about the importance of studying the performance on data groups versus evaluating the whole data altogether where the negative performance tends to average out. Unfortunately, most of the data that recommenders use are missing important features like user information. That's why forming meaningful slices, such as those proposed in \cite{chung2019slice,chung2019automated}, work in theory but can be extremely difficult to apply on the common datasets for interactive systems. Our method overcomes this for the first most common types of datasets in such systems which are the rating-based datasets.

In general, most of the benchmarking toolkit proposals do not include studies about the quality of the data used for the recommender application, nor provide a detailed coverage of the concept related to analyzing performance on smaller groups. The neighborhood-based method proposed in this study fills this gap and provides a new layer of evaluation that better tracks the performance of models across different groups.

\section{Conclusion}
Evaluating the effectiveness of recommendation systems and testing their performance as the underlying data evolves with time remains a very tough challenge that is yet to be tackled. The current evaluation state of the art has seen some improvement in the benchmarking field where researchers were able to create several evaluation frameworks \cite{sun2020we,anelli2021elliot,ovaisi2022rgrecsys}, in an attempt to standardize the assessment process. However, the concept of reporting the performance of models on dataset subgroups hasn't been covered for recommenders, knowing that it is a crucial. As pointed out throughout this work, several studies have discussed the importance of evaluation on smaller data groups, each from a different perspective. For example, dataset natural noise \cite{jurdi2021critique}, the Simpson's paradox \cite{macdonald2021simpson}, and fairness on the user level \cite{singh2018fairness}. In this research, we proposed a new evaluation mechanism for recommender systems called neighborhood-based evaluation that can be used as an assessment tool to track the performance of a recommender on certain portions of the data. The method works on all rating-based datasets and was tested in a systematic evaluation approach on two primary categories of recommenders: prediction optimized models and ranking optimized models. Neighborhoods are created through a KNN clustering method and pass through a series of testing to determine which of them is \textit{critical} within the context of a recommender. The results showed that various neighborhoods performed in a different manner on critical neighborhoods and proved how the normal evaluation on the full datasets can hide the bad performance results on smaller groups. The neighborhood-based evaluation method is a modular framework that acts as an additional layer of evaluation to identify and track the performance on more compact groups of the system. This opens up the possibility of utilizing the neighborhood mechanism for several applications like the implementing a hybrid recommender model or report on the fairness of a certain algorithm within a certain environment. The potential problems we set to address in a future work include extending the neighborhood method to other dataset types that have different features and scale the clustering approach to include different grouping methods.

\section{Acknowledgments}
This work has been supported by the Lebanese University Research Program.

\bibliographystyle{ACM-Reference-Format}
\bibliography{manuscript}

\appendix

\end{document}